\documentclass[conference]{IEEEtran}
\IEEEoverridecommandlockouts
\usepackage{cite}
\usepackage{amsmath,amssymb,amsfonts, multicol}
\usepackage{algorithmic}
\usepackage{graphicx}
\usepackage{textcomp}
\usepackage{url}
\usepackage{lipsum}
  
\usepackage{subfigure}
\usepackage{xcolor}
\def\BibTeX{{\rm B\kern-.05em{\sc i\kern-.025em b}\kern-.08em
    T\kern-.1667em\lower.7ex\hbox{E}\kern-.125emX}}
\begin{document}

\title{Estimation of lunar surface dielectric constant using MiniRF SAR data\\
}

\author{\IEEEauthorblockN{Nidhi Verma, \textit{Student Member, IEEE}, Pooja Mishra, \textit{Member, IEEE}, Neetesh Purohit, \textit{Member IEEE}}}

\maketitle

\begin{abstract}
A new model has been developed to estimate dielectric constant ($ \epsilon $) of the lunar surface using Synthetic Aperture Radar (SAR) data. Continuous investigation on the dielectric constant of the lunar surface is a high priority task due to future lunar mission’s goals and possible exploration of human outposts. For this purpose, derived anisotropy and backscattering coefficients of SAR images are used. The SAR images are obtained from Miniature Radio Frequency (MiniRF) radar onboard Lunar Reconnaissance Orbiter (LRO). These images are available in the form of Stokes parameters, which are used to derive the coherency matrix. Derived coherency matrix is further represented in terms of particle anisotropy. This coherency matrix’s elements compared with Cloud’s coherency matrix, which results in the new relationship between particle anisotropy and coherency matrix elements (backscattering coefficients). Following this, estimated anisotropy is used to determine the dielectric constant. Our model estimates the dielectric constant of the lunar surface without parallax error. The produce results are also comparable with the earlier estimate. As an advantageous, our method estimates the dielectric constant without any apriori information about the density or composition of lunar surface materials. The proposed approach can also be useful for determining the dielectric properties of Mars and other celestial bodies.  
\end{abstract}

\begin{IEEEkeywords}
MiniRF data, dielectric constant, coherency matrix, particle anisotropy.     
\end{IEEEkeywords}

\section{Introduction}
The continuous examination of the lunar surface dielectric constant is a high priority task for many countries due to the ongoing, future lunar missions and exploration of human outposts \cite{1, 2, 3, 4, 5}. Some of these missions are Chandrayaan-3, Smart Lander for Investigating Moon (SLIM), and Spacebit mission (first UK robotic lunar mission), etc. The planning of these missions is to land their rover on the lunar surface to explore resources. For landing the rover, these missions require the knowledge of the regolith materials \cite{6, 7}. The selection of landing sites depends on the particular mission objectives. One of the common goals is the analysis of the lunar surface's material properties, which includes the estimation of lunar surface dielectric constant \cite{5, 8, 9}. 
\par In past number of methods have been developed for estimating lunar surface dielectric constant \cite{8, 10,11, olhoeft1975dielectric_12, wang2010lunar_13, yushkov2019modeling_14}.The dielectric constant estimation is not a new priority, but it is important from several years \cite{10, 11}. In 1975 an empirical relationship between the dielectric constant and bulk density had been derived by Olhoeft and Strangway using lunar samples returned by the Apollo missions \cite{olhoeft1975dielectric_12}. During the Apollo missions 11, 12, 13, 14, 15, 16, and 17, various samples were taken from the lunar regolith. There were seven samples of such kind. Four of the 10084 (Apollo 11), 12070 (Apollo 12), 14163 (Apollo 14), 66041 (Apollo 16)  are regolith samples \cite{1969_15, Pollack_1972_17}. Dielectric results from laboratory experiments revealed the nature of lunar regolith. Another method in 1998 was given by Chyba et al. in terms of energy loss during propagation in an absorbing medium\cite{Chyba_1998_16}. Alan et al. \cite{10}  and Pollack et al. \cite{Pollack_1972_17} analyzed the relationship between the dielectric constant and density of large regions using radar echoes. Campbell gave the formula for estimation of dielectric constant in terms of radar backscattering coefficients. Avik et al. \cite{8} gave another method of dielectric constant estimation. Hagfor’s observations show a dielectric constant of 2.7 for the less dense upper regolith \cite{Hagfors_1970_18}. The overall dielectric constant of the Moon at the centimeter wavelength is 2.7. This is due to a mixture in which 95\% of the material consists of fine grain (dust) material \cite{thompson2014specular_19}.
\par Hence, the dielectric constant ($\epsilon$) estimation model has been developed as per the current research requirement for future lunar studies using Miniature Radio Frequency (MiniRF) SAR data onboard Lunar Reconnaissance Orbiter (LRO) \cite{Raney_2011_20}. Our model has applied to MiniRF SAR data of the Apollo 17 landing site Taurus-Littrow valley and Sinus Iridium area. The avaible MiniRF data is not parallaxe free which can affect the results from SAR images \cite{fa2018unravelling, virkki2019modeling}. For removing the parallax error MiniRF SAR raw data has been process using ISIS3 open source software. Further, Stokes parameters has been derived \cite{Raney_2011_20}. These Stokes parameters have been further used to derive the coherency matrix, which is reflection symmetric. The derived coherency matrix is further converted into particle anisotropy form. Further, derived coherency has been compared with the coherency matrix given by Cloude \cite{Cloude_2012_21}. This comparison provides a unique relation between Stokes parameters and particle anisotropy. Finally, the model has been developed for the estimation of dielectric constant in terms of particle anisotropy.
\par The developed model for lunar surface dielectric constant measurement has the advantage over laborite’s measurement dielectric constant. For the estimation of dielectric constant, density extraction is required; however, in our work, such kind of apriori information is not needed \cite{yushkov2019modeling_14, Palmer_2015_22, Fa_2011_23}. Here, we have used the advantage of microwave scattering information for estimating dielectric constant. Our method can also be relevant for estimating the dielectric constant of other celestial bodies such as Mars etc.

\section{Data and Study Site Details}
In this work, hybrid polarimetric S-band MiniRF data has been used. This data is obtained by transmitting (Left/right) circular polarization and receiving horizontal and vertical polarization coherently. The available MiniRF data level-2 in PDS is not orthorectified. Due to the topographic varaions the parallax error comes into the existance \cite{fa2018unravelling, virkki2019modeling}in the MiniRF data. Some times the SAR data produces 20\% or higher like 50-60\% (mostly in case of opposite sens (OC) and same sens (SC) polarization)\cite{fa2018unravelling, virkki2019modeling}. This error effect the results in the previous developed method for estimation lunar surface dielectric constant. Hence, level-1 raw data again reprocess and parallax error has been removed using the process given by Fa et al.\cite{fa2018unravelling}. This data is further used for the extration of Stokes parameters ($S_1$, $S_2$, $S_2$ and $S_3$). The selected study areas for this paper are Taurus-Littrow valley and  Sinus Iridium., the $S_1$ (total power) images of which are shown in Figure 1 (a) and Figure 1 (b) (before parallax correction), respectively. The complete description of these images is given in Table 1. We have chosen these areas due to the earlier study suggested that they are more suitable for landing purposes and less affected by roughness. 

\par The mass wasting is process which usally present on lunar surface. This mass wasting some time effect the SAR image due to high roughness. Hence further to avoid the missinterpreation from SAR images the GDL DTM 100M was used to mask the high slope area (slope $<$ 5$^{\circ}$) after parallex correction.
\begin{figure*}[t]
	\centering
	\includegraphics[width=.7\linewidth]{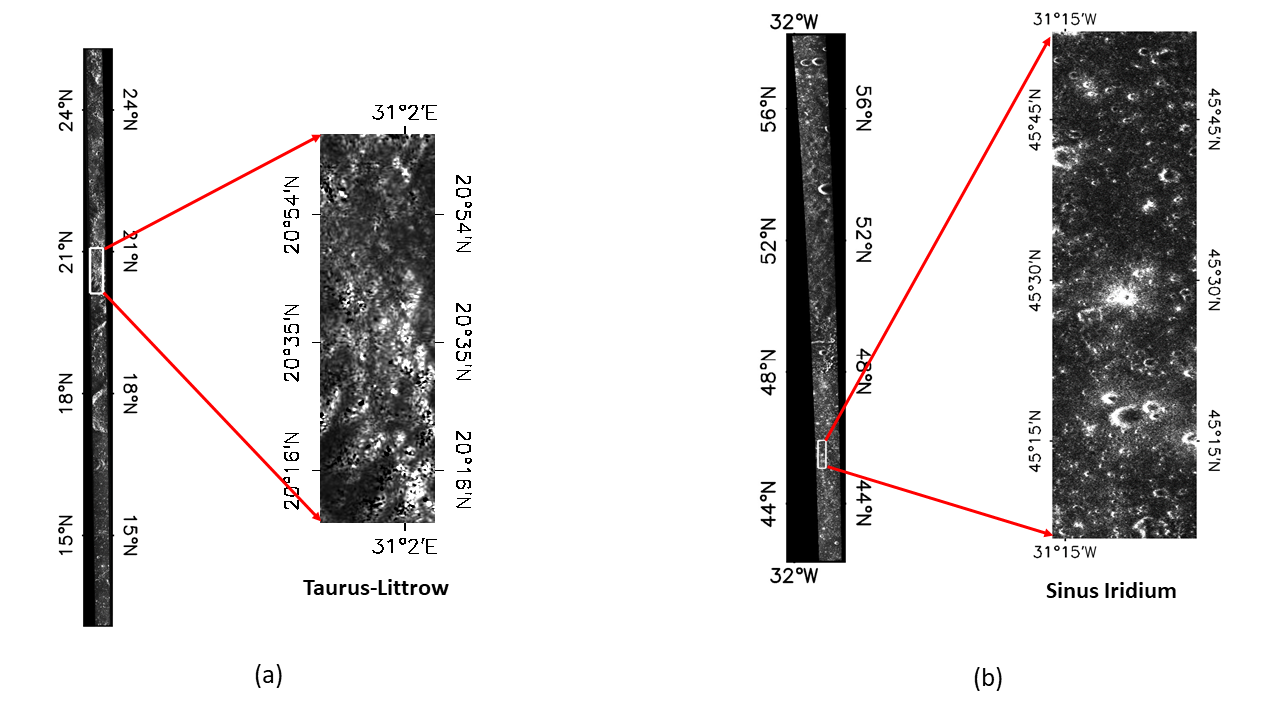}
	\caption{(a) Total power image ($S_1$)of Taurus–Littrow landing area; (b) Total power image of ($S_1$) Sinus-Iridum area.}
	\label{}
\end{figure*}

\begin{table}[htp]
	\caption{Data description} 
	\centering 
	\begin{tabular}{c c c} 
		\hline\hline 
		Parameters & Taurus-Littrow
		valley & Sinus Iridium \\ [0.5ex] 
		\hline 
		Central frequency & 2.38 GHz& 2.38 GHz\\
		Incidence angle & $52.21^{\circ}$& $53.73^{\circ}$\\      
		Center longitude & 	$30.586469^{\circ}$& $34.821143^{\circ}$\\
		Center latitude & 	$17.234865^{\circ}$& $41.744494^{\circ}$\\
		Maximum latitude &	$25.33734^{\circ}$& $45.443823^{\circ}$\\
		Minimum latitude&	$9.146916^{\circ}$& $38.046091^{\circ}$\\
		Westernmost longitude&	$9.146916^{\circ}$& $324.67553^{\circ}$\\
		Easternmost longitude&	$30.154224^{\circ}$& $325.622395^{\circ}$\\
		\hline  
	\end{tabular}
\end{table}

\section{Methodology}
In this paper, we have developed the model for estimating the dielectric constant ($\epsilon$) of the lunar surface using MiniRF Synthetic Aperture Radar (SAR) data by using the concept of particle anisotropy concept. Our model assumes that the surface regolith particles are the same as the cloud of randomly oriented particles\cite{Thompson_2011_24}. This assumption states that the regolith particles size is smaller than the S-band wavelength, which is used for the volume scattering mechanism from the ‘‘surface regolith” \cite{Thompson_2011_24}.The steps for the development of the proposed model are as follows

\subsection{Generation of scattering matrix }
The scattering matrix is defined below in equation (1) using wave coordinate Back Scattering Alignment  (BSA) in terms of particles anisotropy (Ap)\cite{cloude2010polarisation}

\begin{equation} \label{eu_eqn}
[S] =
\begin{bmatrix}
S\textsubscript{11}& S\textsubscript{12}\\
S\textsubscript{21}& S\textsubscript{22}
\end{bmatrix}=
\begin{cases}
S\textsubscript{11}=(A\textsubscript{p}-1)sin\textsuperscript{2}\theta\cos\textsuperscript{2}\tau+1\\
S\textsubscript{12}=(A\textsubscript{p}-1)cos\theta\cos\textsuperscript{2}\tau\\\
S\textsubscript{21}= (A\textsubscript{p}-1)sin\theta\cos\theta\cos\textsuperscript{2}\tau\\
S\textsubscript{22}=(A\textsubscript{p}-1)cos\textsuperscript{2}\theta\cos\textsuperscript{2}\tau+1\\
\end{cases}  
\end{equation}
\par Where angle $ \theta $ and $ \tau $ represent the rotation and tilt angle of the spheroid, respectively. The $A\textsubscript{p}$ is particle anisotropy, which is also known as particle polarizability. The value of $A\textsubscript{p}$ is defined below in equation (2)
\begin{equation} \label{eu_eqn}
Particle~anisotropy~(A\textsubscript{p} )\geqslant 0		
\end{equation}
\subsection{Generation of Coherency matrix }
The scattering matrix in equation (1) is in complex symmetric form. Hence it can be represented in the form of eigenvalue expansion form as shown in equation (3)\cite{cloude2010polarisation}

\begin{equation} \label{eu_eqn}
[S] =
\begin{bmatrix}
cos\theta&\sin\theta\\
-sin\theta& cos\theta
\end{bmatrix}
\begin{bmatrix}
1&0\\
0& sin\textsuperscript{2}\tau+A\textsubscript{p}sin\textsuperscript{2}\tau
\end{bmatrix}
\begin{bmatrix}
cos\theta&-sin\theta\\
sin\theta& cos\theta
\end{bmatrix}
\end{equation}
The scattering matrix further reformulated into the following scattering vector ($k$) form as given by equation (4)
\begin{equation} \label{eu_eqn}
k =
\begin{bmatrix}
1&0&0\\
0& cos2\theta&-sin2\theta\\
0&sin2\theta&cos2\theta\\
\end{bmatrix}
\begin{bmatrix}
2+Xcos\textsuperscript{2}\tau\\
Xcos\textsuperscript{2}\tau\\
0\\
\end{bmatrix}
\end{equation}
\begin{align*}
Where~X=A\textsubscript{p}-1
\end{align*}

Further, we have computed the Coherency matrix [$T$] $(3\times3)$ using $k$ as shown in equation (5)

 \begin{equation} \label{eu_eqn}
[T] =\dfrac{1}{2}kk\textsubscript{*}\textsuperscript{T}
\end{equation}

\par Where $k^{*}$ represents complex conjugate of $k$, T represents the transpose. 
The final simplified form of coherency matrix is given below \\
\noindent
\begingroup
\begin{equation} \label{eu_eqn}
\resizebox{0.5\textwidth}{!}
{$
	[T] =\dfrac{1}{2}
	\begin{bmatrix}
	|2+Xcos\textsuperscript{2}\tau|^{2}&(2+Xcos\textsuperscript{2}\tau)X^{*}cos\textsuperscript{2}\tau\cos2\theta& (2+Xcos\textsuperscript{2}\tau)X^{*}cos\textsuperscript{2}\tau\sin2\theta\\
	(2+X^{*}cos^{2}\tau)Xcos^{2}\tau\cos2\theta&|Xcos\textsuperscript{2}\tau\cos2\theta|^{2}&X^{2}cos^{4}\tau\cos2\theta\sin2\theta\\
	(2+X^{*}cos^{2}\tau)Xcos^{2}\tau\sin2\theta&X^{*2}cos^{4}\tau\cos2\theta\sin2\theta&|Xcos\textsuperscript{2}\tau\cos2\theta|^{2}\\
	\end{bmatrix}
	$}
\end{equation}
\endgroup\\
Here, we have found that the off-diagonal elements of [T] will be zero for all possible value of $ \tau $ and $ \theta $. However, non-diagonal elements maintain scattering information.
\subsection{Coherency matrix conversion into eigenvalue decomposition matrix form}
Next, equation (6) converted into the Eigenvalue decomposition form with three diagonal values are , $\lambda\textsubscript{11}$, $\lambda\textsubscript{22}$ and  $\lambda\textsubscript{33}$

\begin{equation} \label{eu_eqn}
[T] =
\begin{bmatrix}
1&0&0\\
0&1&0\\
0&0&1\\
\end{bmatrix}
\begin{bmatrix}
\lambda\textsubscript{11}&0&0\\
0&\lambda\textsubscript{22}&0\\
0&0&\lambda\textsubscript{33}\\
\end{bmatrix}
\begin{bmatrix}
1&0&0\\
0&1&0\\
0&0&1\\
\end{bmatrix}
\end{equation}

where  $\lambda\textsubscript{11}$, $\lambda\textsubscript{22}$ and  $\lambda\textsubscript{33}$.

\subsection{Calculation  of eigenvalues }
We have averaged the matrix given by equation (7) overall possible angles values of $\theta$ and $\tau$, which provides the probability distribution of the random distribution case and is equated to diagonal elements of equation (8). For this purpose, the probability distributions p(..) for a random distribution are defined as shown in equations (9), (10) given below \cite{Arii_2010_25}
\begin{equation}
p(\theta)=\dfrac{d\theta}{2\pi}
\end{equation}
\begin{align*}
where~
-\pi\le\theta\textless\pi
\end{align*}
\begin{equation}
p(\theta)=\dfrac{cos\tau d\tau}{2}
\end{equation}
\begin{align*}
where~
\dfrac{-\pi}{2}\le\tau\textless\dfrac{\pi}{2}
\end{align*}
From the above equation (7), eigenvalue $\lambda\textsubscript{11}$ ($t\textsubscript{11}$) is represented in explicit form using equation (9) and (10)
\begin{equation}
\lambda\textsubscript{11}=\int\limits_{\dfrac{-\pi}{2}}^{\dfrac{\pi}{2}}\int\limits_{-\pi}^{\pi} (4+4Xcos^{2}\tau+X^{2}cos^{4}\tau)cos\tau d\theta d\tau
\end{equation}
\begin{equation}
\lambda\textsubscript{11}=2+\dfrac{4}{5}X+\dfrac{4}{15}X^{2}
\end{equation}
Here, $\lambda\textsubscript{11}$ is also known as first element of Cohrency matrix.
\subsection{Calculation  of eigenvalues using Cloude approach}
The traditional form of Cohrency matrx is reprsented by euation (12)

\begin{equation} \label{eu_eqn}
[T] =
\begin{bmatrix}
t\textsubscript{11}&t\textsubscript{12}&t\textsubscript{13}\\
t\textsubscript{21}&t\textsubscript{22}&t\textsubscript{23}\\
t\textsubscript{31}&t\textsubscript{32}&t\textsubscript{33}\\
\end{bmatrix}
\end{equation}
\par Where $t\textsubscript{11}, t\textsubscript{12}, t\textsubscript{13}, t\textsubscript{21}, t\textsubscript{22}, t\textsubscript{23}, t\textsubscript{31}, t\textsubscript{32}$ and $t\textsubscript{33}$ are elements of the coherency matrix. For computing the coherency matrix elements Cloude approach \cite{Cloude_2012_21} was used to in terms of Stokes parameters as shown in equation (13)  

\begin{equation} \label{eu_eqn}
\begin{bmatrix}
S\textsubscript{1}\\
S\textsubscript{2}\\
S\textsubscript{3}\\
S\textsubscript{4}\\
\end{bmatrix}=
\begin{bmatrix}
\dfrac{1}{2}(t\textsubscript{11}+t\textsubscript{22}+t\textsubscript{33})\pm Img(t\textsubscript{23})\\
Real(t\textsubscript{12})\pm Img(t\textsubscript{13})\\
Real(t\textsubscript{13})\pm Img(t\textsubscript{12})\\
Img(t\textsubscript{23})\pm \dfrac{1}{2}(t\textsubscript{22}+t\textsubscript{33}-t\textsubscript{11})\\
\end{bmatrix}
\end{equation}
\par  Here + sign is used for Left Hand Circular (LHC) polarized signal and – sign is used for Right Hand Circular (RHC) polarized signal \cite{Cloude_2012_21}. From the above equation (13) for RHC polarized signal, we have obtained the equation (14).
\begin{equation}
S\textsubscript{1}+S\textsubscript{4}=t\textsubscript{11}
\end{equation}
Equation (14) can also be known as the opposite sense circular polarization (OC). Further, equation (14) is equated with equation (11), which results in the following equation

\begin{equation}
S\textsubscript{1}+S\textsubscript{4}=2+\dfrac{4}{5}X+\dfrac{4}{15}X^{2}
\end{equation}

\begin{align*}
Where~X=A\textsubscript{p}-1
\end{align*}
\subsection{Calculation  of anisotropy }
At last, anisotropy has been calculated with the help of the following equation
\begin{equation}
A\textsubscript{p}=abs(X+1)
\end{equation}
\subsection{Estimation  of dielectric constant  }

The above equation (15) is used to invert the particle anisotropy ($A_p$). At last dielectric constant inverted using equation (17)
\begin{equation}
Dielectric~constant~{\epsilon}=\dfrac{1}{A\textsubscript{p}}+abs\left(
\dfrac{1}{A_p}-(2A_p-1)\right)
\end{equation}
The dielectric constant ($\epsilon$) lies the limit for particle anisotropy as shown in equation (18) \cite{ablitt2000characterisation}
\begin{equation}
\dfrac{1}{\epsilon}\textless A_p\textless\dfrac{\epsilon+1}{2}
\end{equation}
\subsection{Results and discussions}
The dielectric constant values have estimated using our model for the Taurus–Littrow valley and Sinus Iridium. Further statistical analyses have been performed using a step plot to determine the distribution of these two classes. 

In Figure 2(a) the class 1 shows 97.55\% (mean value is 2.56$\pm$0.16)  pixels are having the value of dielectric constant less than  3, and class 2 shows 2.55\% (mean value is 3.63$\pm$0.54) of pixels having dielectric constant greater than 3 (mean value is 3.62) for Taurus–Littrow valley. The overall mean value of dielectric constant for Taurus–Littrow valley has been found 2.56$\pm$0.25. Similarly, for Sinus-Iridium overall dielectric constant has been found 2.54$\pm$0.43. The distribution shows 96.16\% (mean value is 2.48$\pm$0.14 for class 1) pixels having a dielectric constant value less than 3 and 3.84\% pixels having dielectric constant greater than 3 for (mean value is 4.22$\pm$1.15 for class 2), respectively as shown in Figure 2(b).The results of dielectric constant for Taurus–Littrow valley also has been found closer to previous estimates \cite{11, Hagfors_1970_18}. 
\begin{figure}[]
	\centering	
	\begin{minipage}{.5\textwidth}
		\centering
		\subfigure[Taurus-Littrow valley area.]{%
			\label{fig:first}%
			\includegraphics[height=2in]{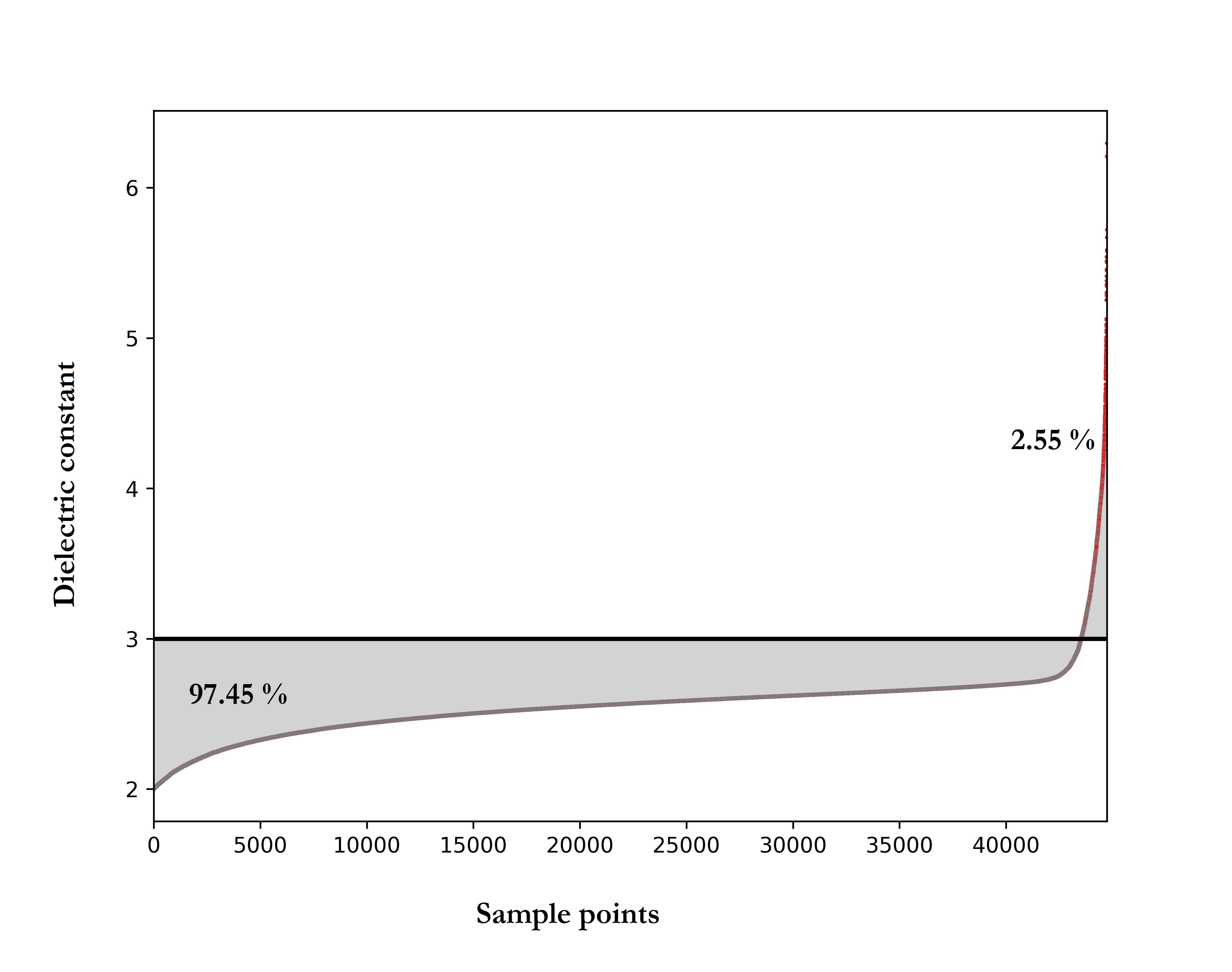}}%
	\end{minipage}\hfill
	\begin{minipage}{.5\textwidth}
		\centering
		\subfigure[Sinus-Iridium area.]{%
			\label{fig:second}%
			\includegraphics[height=2in]{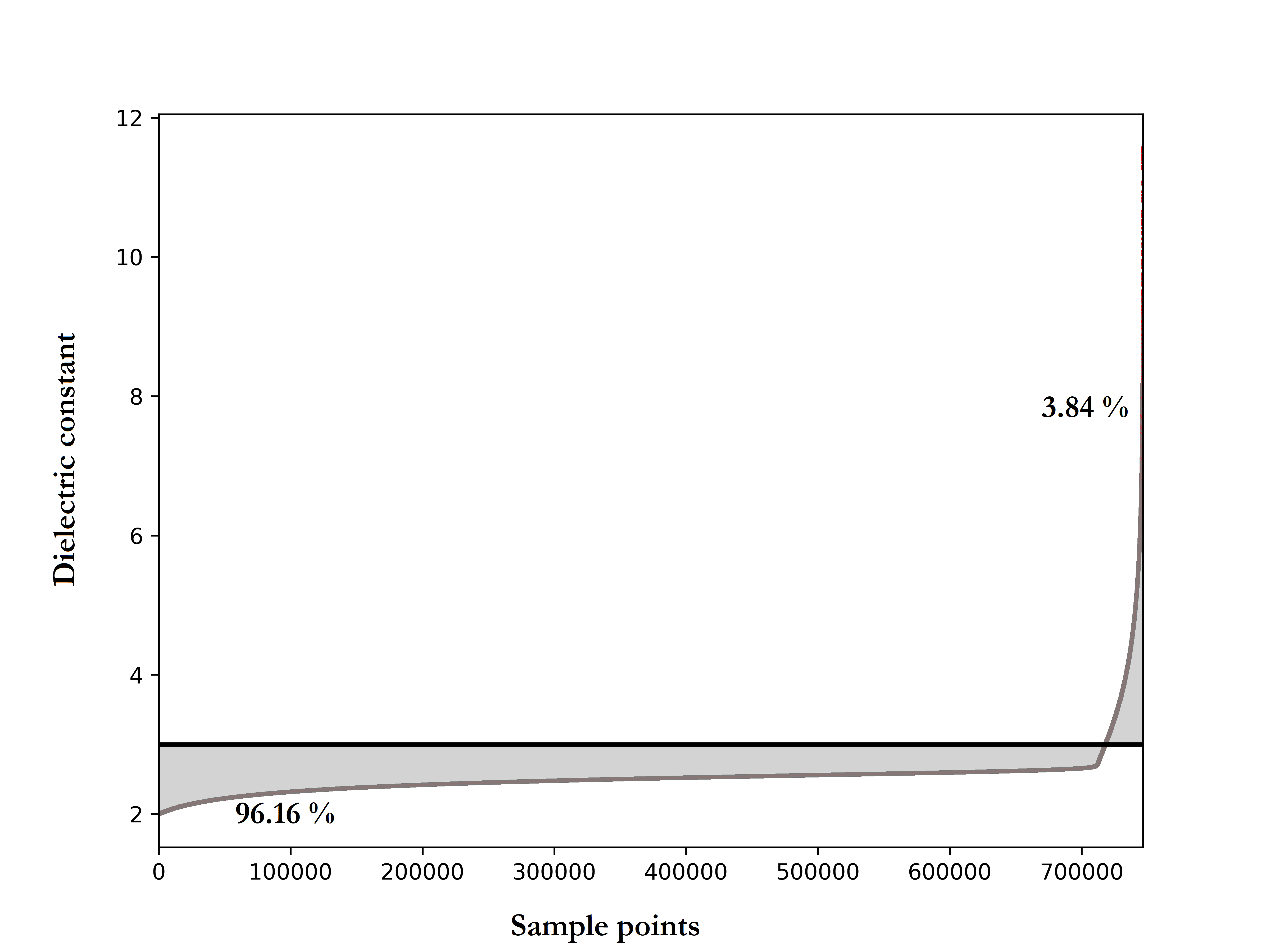}}%
	\end{minipage}
	\caption{(a) Step plot for dielectric constant of Taurus–Littrow valley; (b) Step plot for dielectric constant distribution of Sinus-Iridium.}
\end{figure}

\subsection{Application of developed model}
\noindent The developed model has been applied on the Chandrayaan-2 alternate landing site MiniRF SAR (S-band) image with center latitude and longitude -66.077164$^\circ$, -18.168836$^\circ$. Chandrayaan -2 landing site total power image ($S_1$) is given in Figure 3 (a) (before parallax correction). After parallax correction the overall mean value of dielectric has been found 2.52$\pm$0.27.The distribution of dielectric constant is shown in Figure 3 (b) with 97.75\% (mean value is 2.47$\pm$0.14) and 2.27\% (mean value is 3.73$\pm$0.72). These results are also similar to the Taurus–Littrow valley area. 
\begin{figure}[]
	\centering	
	\begin{minipage}{.5\textwidth}
		\centering
		\subfigure[Chandrayaan-2 alternate site.]{%
			\label{fig:first}%
			\includegraphics[height=2in]{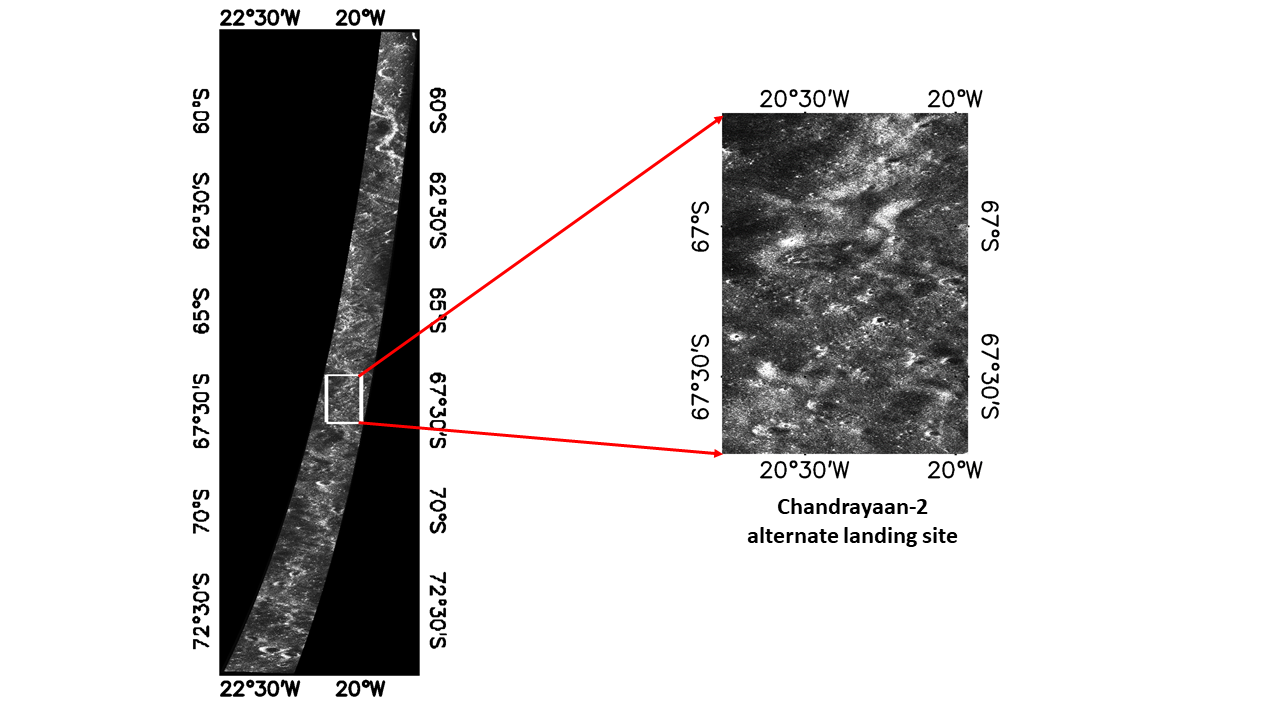}}%
	\end{minipage}\hfill
	\begin{minipage}{.5\textwidth}
		\centering
	\end{minipage}\hfill
	\begin{minipage}{.5\textwidth}
		\centering
		\subfigure[Dielectric constant of Chandrayaan-2 alternate site.]{%
			\label{fig:second}%
			\includegraphics[height=2in]{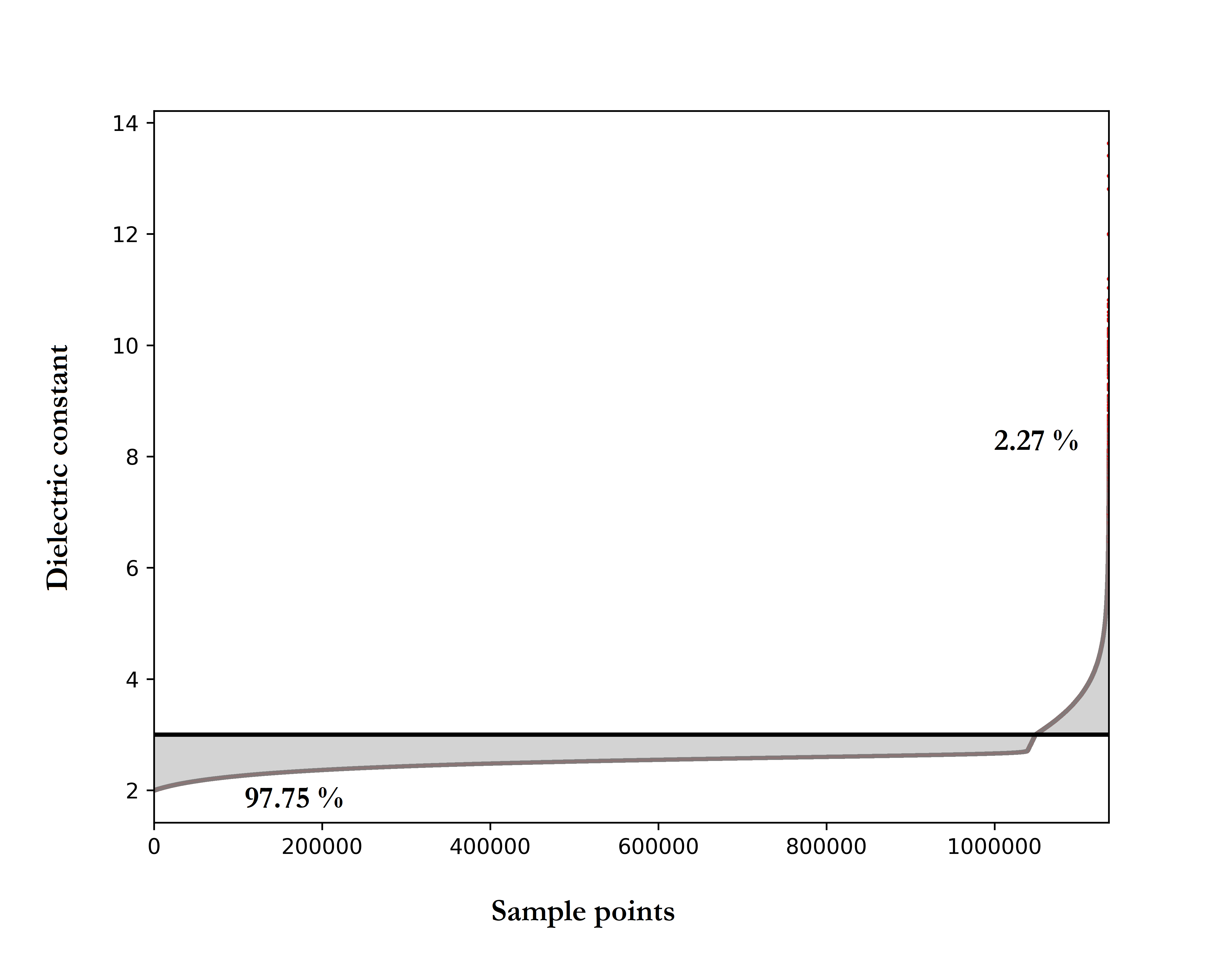}}%
	\end{minipage}\hfill
	
	\caption{(a) Chandrayaan-2 alternate site.;  (b) Step plot for dielectric constant distribution.}
\end{figure}
We have constructed a mathematical model to estimate the lunar surface dielectric constant. The important relation between particle anisotropy and Stokes parameters has been developed. This requires a Coherency matrix, which is generated using MiniRF SAR data. Using this model, we estimate a dielectric constant parameter to describe the lunar regolith's electromagnetic property. Analysis of the different lunar landing sites data on S-band frequency implies substantial contamination of the fine dust materials. The results from dielectric constant provides less bias than other methods and proved to be feasible in theory and comparable with earlier estimates. These observations provide a foundation for microwave SAR data can be useful for implementing our model in the future for other celestial bodies.

\section{Conclusion}


The new method has been developed for the estimation of lunar surface dielectric constant with hybrid polarimetric MiniRF SAR data. The results from the proposed method for Taurus-Littrow valley and Sinus Iridium area found to be comparable with the earlier estimates. The proposed method for dielectric constant estimation has several advantages, such as, is easier to implement and no aprior information, free from parallax error and no need of auxiliary data are needed, such as surface material composition like bulk density and so on. It is more suitable for the MiniRF or MiniSAR data collected by SAR instruments. The original intention of this method is only for the dielectric estimation of the lunar surface. The estimated dielectric constant can be futher useful for the extraction of loss tangent, bulk density and penetration depth.

\section*{Acknowledgment}
~The authors would like to thank The Geosciences Node of NASA's Planetary Data System (PDS) archives and distribute digital MiniRF SAR data related to the lunar surface study. They would also like to gratify the Ministry of Human Resource and Development, Government of India, for providing the research scholarship for carrying out this work.

\bibliographystyle{IEEEtran}
\bibliography{nidhi}

\end{document}